\documentclass[12pt]{article}
\title{
Chiral Lagrangian with confinement from the QCD Lagrangian}
\author{Yu.A.Simonov \\
Jefferson Laboratory,Newport News,VA 23606,USA ,\\
State Research Center,\\ Institute of Theoretical and Experimental
Physics, Moscow, Russia}
  \newcommand{\be}{\begin{equation}}
\newcommand{\ee}{\end{equation}}  
\def\fun#1#2{\lower3.6pt\vbox{\baselineskip0pt\lineskip.9pt
\ialign{$\mathsurround=0pt#1\hfil ##\hfil$\crcr#2\crcr\sim\crcr}}}
\newcommand{\vex}{\mbox{\boldmath${\rm x}$}}

\newcommand{\veu}{\mbox{\boldmath${\rm u}$}}

\newcommand{\vek}{\mbox{\boldmath${\rm k}$}}
\newcommand{\vey}{\mbox{\boldmath${\rm y}$}}

\newcommand{\vep}{\mbox{\boldmath${\rm p}$}}

\newcommand{\veal}{\mbox{\boldmath${\rm \alpha}$}}
\newcommand{\vegam}{\mbox{\boldmath${\rm \gamma}$}}

\newcommand{\ve}{\vec}
 \newcommand{\lan}{\langle}
 \newcommand{\ran}{\rangle}
\begin{document}

\maketitle

\begin{abstract}
An effective Lagrangian for the light quark in the field of a
static source is derived systematically using the exact field
correlator expansion. The lowest Gaussian term is bosonized using
nonlocal colorless bosonic fields and a general structure of
effective chiral Lagrangian is obtained containing all set of
fields. The new and crucial result is that the condensation of
scalar isoscalar field which is a usual onset of chiral symmetry
breaking and is constant in space-time, assumes here the form of
the confining string and contributes to the confining potential,
 while the rest bosonic fields describe mesons with the $q\bar q$
  quark structure and pseudoscalars play the role of Nambu-Goldstone
fields. Using derivative expansion the effective chiral Lagrangian
is deduced containing both confinement and chiral effects for
heavy-light mesons. The pseudovector quark coupling constant is
computed to be exactly unity in the local limit,in agreement with
earlier large $N_c$ arguments.
\end{abstract}

\section{Introduction}

It was understood long ago \cite{1} that  chiral symmetry breaking
in QCD is responsible for the low mass of pions and therefore the
low-energy limit of QCD can be adequately described by the
effective chiral Lagrangians \cite{2}.

In this approach the Nambu-Goldstone particles are described by
local  field variables and the resulting Effective Chiral
Lagrangian (ECL) is local.

The most general and practically useful form of chiral Lagrangian
was given in \cite{3} and contains  around ten phenomenological
parameters (14 to the 4th order in  $p$), to be found from
experiment.

Being successful in describing low-energy processes with
Nabmu-Goldstone mesons, ECL has two major defects: Firstly, it
does not take  into account the quark  structure of mesons, and
consequently e.g. the formfactor computed in ECL can display in the
meson only mesonic degrees of freedom.

Secondly, ECL completely disregards the phenomenon of confinement,
which is also important at small energies, and hence degrees of
freedom of vacuum gluons, creating confining string are not taken
into  account.

There have been attempts to cure the first defect, namely the
model Lagrangians have been suggested which take into account both
quarks and mesons \cite{4,5}. In  particular the instanton model
of the QCD vacuum has been used to derive the ECL and the Quark-
Meson Lagrangian(QML) \cite{6}. As a result interesting
interconnections of quark and chiral degrees of freedom have been
demonstrated in the example of the nucleon \cite{7}.

However instantons are suppressed in the realistic  vacuum of QCD
\cite{8,9} and moreover the internal consistency of the  instanton
vacuum without confinement  is seriously questioned \cite{10,11}.

Moreover instantonic vacuum lacks confinement and  therefore
cannot cure the second defect of ECL and QML, therefore model ECL
and QML obtained in \cite{6,7} disregard confinement fully.

Finally, the Effective  Quark Lagrangian (EQL)  obtained after
averaging  over gluon degrees  of freedom, contains in principle
infinite number of terms  with growing number of quark fields. It
was shown in \cite{12} that specifically for instanton vacuum all
higher terms are of the same order, while only the lowest term
(so-called 'tHooft Lagrangian) is taken in  the standard instanton
lore.

It is the purpose of the present paper to start a new and
systematic approach to the derivation of QML and ECL from the
first principles - the QCD Lagrangian. In doing so the most
important is to keep gauge invariance at every stage, therefore we
shall consider the simplest gauge-invariant system of a light
guark in the field of a static  antiquark (generalization to only
light quark systems will be done later).

After averaging over gluon fields one obtains EQL with infinite
number of terms,  containing as kernels irreducible gluon Field
Correlators (FC). Recent measurements for the realistic QCD vacuum
 have shown that FC create a hierarchy, where the lowest
(Gaussian) correlator is  dominating \cite{8,9}, while the next
(4th order) correlator contributes around 1\%  to the static
$Q\bar Q$ interaction (note that this  situation is  drastically
different from that of instantonic vacuum, where higher
correlators are equally important). We assume that a similar
hierarchy should be present also in our problem of      a
light-heavy quark, which allows us to study  the resulting EQL
term by term, paying most attention to the lowest, $4q$-piece.
The next step is the bosonization procedure, i.e. an identical
transformation introducing nonlocal colorless bosonic fields,
having different Lorentz and flavour indices due to the use of
Fierz transformation.

A special attention  deserves the $6q$ term, where bosonization
may be done introducing bosonic fields for $\bar q q$ combination
or else introducing baryonic fields for $3q$ combination.

As a result one obtains QML or Quark-Baryon Lagrangian (QBL) in
the most general and rigorous form. These questions  will be
considered in a separate  publication \cite{13}.

As the next step one can use the stationary point analysis to
obtain nonlinear equations for effective bosonic fields.

This is done in a rigorous way and the resulting equations contain
for the scalar-isoscalar part the same equation as was derived
previously \cite{14}  in a gauge-invariant Dyson-Schwinger approach
 not using bosonization.  The careful study of that equation in
\cite{14}-\cite{16} has shown that it describes in the light-heavy
system both confinement and chriral symmetry breaking, which
coexist. This fact means that confinement in the language of
effective meson fields, enters in the  form of condensate of the
scalar field inside the string, while other fields describe
subdominant features of the $q\bar Q$ dynamics.

The paper is organized as follows. In the next section the
averaging over vacuum gluonic fields is done and EQL is obtained.
Special attention is devoted to the gauge invariance and parallel
transporters necessary to ensure it for nonlocal $q\bar q$
combinations. In section 3 the bosonization procedure is done and
Fierz transformation is introduced to obtain final QML with proper
classification in  Lorentz and flavour indices.

In section 4 the resulting ECL is obtained and the stationary
point equations are derived and compared to the previously
obtained in  the Dyson-Schwinger approach. In section 5 the
derivative expansion of the nonlocal bosonized Lagrangian is done
and nonlocal forms of the lowest 2nd and 4th order terms in this
expansion are obtained. Keeping only pion field in addition to
quarks, one derives in section 6 the effective Lagrangian which
appears to have the expected form with the pseudovector quark-pion
coupling constant $g^q_A$. The latter is equal exactly to one, in
agreement with earlier large $N_c$ argument, when local
approximation is made. The concluding section is devoted to the
discussion of confinement and chiral properties of the resulting
ECL in the heavy-light meson case.

\section{The Effective Quark Lagrangian}

Consider the QCD partition function in the Euclidean space-time
\be
Z=\int DAD\psi D\psi^+ e^{-S_0(A)+\int~^f\psi^+(i\hat
\partial+im+g\hat A)~^f\psi d^4x}
\label{1}\ee where $S_0(A)=\frac{1}{4}\int(F^a_{\mu\nu}(x))^2
d^4x$, $m$ is the current quark mass, and the quark operator
$^f\psi_{a\alpha}(x)$ has flavor index $f(f=1,... n_f)$,  color
index $a(a=1, ... N_c)$ and Lorenz bispinor index
$\alpha(\alpha=1,2,3,4)$.

The next step is to integrate over $DA_\mu$ with the weight
$S_0(A)$,  this is the gluon vacuum averaging which
 is denoted by $\lan~~\ran_A$. Before doing this  however one should choose
  the gauge-invariant
system and using an appropriate gauge, express $A_\mu$ through
the field-strength operator $F_{\mu\nu}$
 which would finally appear in
  gauge-invariant combinations  -- Field Correlators (FC)(see \cite{17}
 for review and more  discussion), namely
\be
g^n\lan
F_{\mu_1\nu_1}(x_1)\Phi_{C_1}(x_1,x_2)F_{\mu_2\nu_2}(x_2)\Phi_{C_2}(x_2,x_3),...
F_{\mu_n\nu_n}(x_n)\Phi C_n(x_n,x_1)\ran_A\equiv \Delta^{(n)}
\label{2} \ee where parallel transporters are defined as
\be
\Phi_C(x,y) = P\exp (ig\int_{C(x,y)} A_\mu dz_\mu)\label{3}\ee and
the open contour $C(x,y)$ connects points $x$ and $y$ and can be
arbitrary otherwise.

To achieve this goal one can use the so-called contour gauge
\cite{18,19} which is especially
 convenient in the case of light quark moving in the field of a static antiquark.
 One has for the contour $z_\mu(s,x)$ starting at point $x$ and ending at
$Y=z(0,x)$
\be
A_\mu(x)=\int^1_0 ds\frac{\partial z_\nu(s,x)}{\partial
s}\frac{\partial z_\rho(s,x)}{\partial x_\mu} F_{\nu\rho}
(z(s))\equiv \int^x_Y d \Gamma_{\mu\nu\rho} (z) F_{\nu\rho} (z)
.\label{4} \ee

In particular case,
 when the contour $z_\mu(s,x)$ goes along the shortest (straight) way to the $x_4$- axis and
 then along this axis to some point $Y$,
 which can be at - $\infty$, one has the so-called modified
Fock-Schwinger gauge \cite{20}, which was extensivelly used in
\cite{14} to get EQL. Here one has
\be
A_\mu(\vex,x_4)=\int^{x_i}_0\alpha_\mu(u)
du_iF_{i\mu}(u,x_4),\label{5}\ee and $\alpha_4(u)\equiv 1$ while
for $\mu=1,2,3,$ $\alpha_\mu$ is equal to $$
\alpha(u)=\frac{u_i}{x_i},i=1,2,3. $$

It is convenient to write all expressions in a gauge-invariant
way, using the property\cite{19}, that $\Phi_C$ given by (\ref{3})
is
 identically  equal to unity when the contour $C(x,y)$ lies on $z_\mu(s,x)
$ or $z_\mu(s,y)$. Therefore one can define gauge-covariant
operators referred to the point $Y$,

$$
\psi^{(Y)}(x)= \Phi_{C(x,Y)}\psi(x)\equiv \Phi(Y,x)\psi(x)
$$
$$
\psi^{+(Y)}(x)= \psi^{+}(x)\Phi(x,Y)
$$
\be
F^{(Y)}_{\mu\nu}(x)=\Phi(Y,x) F_{\mu\nu}(x)\Phi(x,Y).
\label{6}
\ee
Here the contour $C(x,Y) $ in $\phi(Y,x)$ goes along $z(s,X)$ from $Y$ to $x$, and in opposite direction in
$\Phi(x,Y)$.

For the field correlators referred to the same point $Y$ one can
write ($ab,cd$-- fundamental color indices, Lorentz indices are
suppressed for
 simplicity reasons)

\be
g^2\lan(F^{(Y)} (x))_{ab}(F^{(Y)}(y))_{cd}\ran=
\frac{\delta_{ad} \delta_{bc}}{N_c^2} g^2\lan tr (F^{(Y)}(x)
F^{(Y)}(y) )\ran \label{7} \ee
 $$ g^3\lan(F^{(Y)}(x))_{ab}
(F^{Y}(y))_{cd} (F^{(Y)}(z))_{ef}\ran= \delta_{bc}\delta_{de}
\delta_{af} \left[-\frac{g^3 tr\lan xzy\ran}{
N_c(N_c^4-1)}+\right. $$
\be
\left.+ \frac{N_c}{N_c^4-1} g^3 tr \lan xyz\ran \right] + \delta_{be} \delta_{fc} \delta_{ad}
\left[-\frac{g^3tr\lan xyz\ran}{N_c(N_c^4-1)} +
\frac{N_cg^3 tr\lan xzy\ran}{N_c^4-1}\right]\label{8}
\ee
where notation is introduced e.g. $tr\lan xyz\ran \equiv
\lan tr (F^{(Y)} (x) F^{(Y)}(y) F^{(Y)}(z))\ran.$

Derivation of (\ref{7}),(\ref{8}) is given in Appendix 1.

Let us first concentrate on the bilocal correlator (\ref{7}). From
(\ref{4}), (\ref{5}) it is clear that in the average value of
$\lan A_\mu(x) A_\nu(y)\ran$ the  arguments of $ F(z(s,x))$ and
$F(z(s,y))$ are separated by the distance $r\sim T_g$, where $T_g$
is the gluonic correlation length \cite{21}, which was measured on
the lattice \cite{22} and estimated analytically \cite{23} to be
$T_g\sim 0.2 $fm (or even smaller, if data on gluelump masses
\cite{24} are used). For such distances $r\sim T_g$, which satisfy
$r\ll |x-Y|, |y-Y|$ or for the gauge (\ref{5}), $r\ll |\vex|,
|\vey|,$ one  has an estimate
\be
\lan tr F^{(Y)}(x)F^{(Y)}(y)\ran= \lan tr (F(x)\Phi(x,y)
F(y)\Phi(y,x))\ran+
O(\frac{r^2}{\vex^2},\frac{r^2}{\vey^2})\label{9} \ee
 where the correlator on the r.h.s. of (\ref{9})
is connected by straight lines from $x$ to $y$. A similar estimate
holds for the triple FC (\ref{8}), and in what follows we shall
use the straight-line form (\ref{9}) which is independent on the
position of the reference point $Y$.

For that correlator one can use the general representation found
in \cite{25}
\be
\frac{g^2}{N_c}\lan tr F_{\mu\nu} (x) \Phi(x,y) F_{\rho\sigma} (y)
\Phi (y,x)\ran= D(x-y) (\delta_{\mu\rho} \delta_{\nu\sigma}-
\delta_{\mu\sigma} \delta_{\nu\rho})
+\Delta^{(1)}_{\mu\nu,\rho\sigma}\label{10} \ee where
$\Delta^{(1)}$ has the structure of  a full derivative and
therefore does not contribute to confinement, its
 nonperturbative (NP) part is much smaller than that of $D$ and
 we shall omit it.

The function $D(u)$ has NP part which was measured in \cite{22}
and has exponential form, \be D(u)
=D(0)\exp(-\frac{|u|}{T_g}).\label{11} \ee

Finally, the string tension $\sigma$ can be expressed through
$D(u)$ ( and higher correlators) and at least for static quarks
$D(u)$ yields dominant (up to few
 percent \cite{8,9}) contribution to $\sigma$,
\be
\sigma=\frac12 \int D(u)d^2u. \label{12} \ee Having in mind all
the above relations, one can now average over gluonic fields in
(\ref{1}) to obtain
\be
Z=\int D\psi D\psi^+ e^{\int~^f\psi^+(i\hat \partial+im)^f\psi
d^4x} e^{L^{(2)}_{EQL}+L^{(3)}_{EQL}+...}\label{13} \ee where the
EQL proportional to $\lan\lan A^n\ran\ran$ is denoted by
$L_{EQL}^{(n)}$, \be L^{(2)}_{EQL}=\frac{g^2}{2}\int
d^4xd^4y~^f\psi^+_{a\alpha}(x)
~^f\psi_{b\beta}(x)~^g\psi^+_{c\gamma}(y)~^g\psi_{d\varepsilon}(y)
\lan  A^{(\mu)}_{ab}(x) A^{(\nu)}_{cd}(y)\ran
\gamma^{(\mu)}_{\alpha\beta} \gamma^{(\nu)}_{\gamma\varepsilon}
\label{14} \ee $$ L_{EQL}^{(3)}=\frac{g^3}{3!}\int
d^4xd^4yd^4z~^f\psi^+_{a\alpha}(x)
~^f\psi_{b\beta}(x)~^g\psi^+_{c\gamma}(y)~^g\psi_{d\varepsilon}(y)~^h
\psi^+_{e\rho}(z)~^h\psi_{f\sigma}(z)\times $$
\be
\times\lan A^{(\mu)}_{ab}(x)A^{(\nu)}_{cd}(y)
A^{(\lambda)}_{ef}(z)\ran
\gamma^{(\mu)}_{\alpha\beta}\gamma^{(\nu)}_{\gamma\varepsilon}
\gamma^{(\lambda)}_{\rho\sigma}\label{15} \ee

Average of gluonic fields can be computed using
(\ref{5}),(\ref{7}), (\ref{8}) as
\be
g^2\lan A^{(\mu)}_{ab}(x) A^{(\nu)}_{cd}(y)\ran=
\frac{\delta_{bc}\delta_{ad}}{N_c} \int^x_0
du_i\alpha_{\mu}(u)\int^y_0 dv_k\alpha_\nu(v)
D(u-v)(\delta_{\mu\nu}\delta_{ik}-\delta_{i\nu}\delta_{k\mu}).
\label{16} \ee

The corresponding expression for the triple average $\lan A^3\ran$
is given in Appendix 1.

As it was argued in \cite{26} the dominant contribution at large
distances from
 the static antiquark is given by the color-electric fields, therefore we shall write down
 explicitly $L_{EQL}^{(2)} (el)$ for this case, i.e. taking $\mu=\nu=4$ in (\ref{14}),
(\ref{16}), while the general case, also for generalized gauge
(\ref{4})
 is given in Appendix 2. As a result one has
\be
L_{EQL}^{(2)}(el)=\frac{1}{2N_c}\int d^4x\int
d^4y~^f\psi^+_{a\alpha}(x)~^f\psi_{b\beta}(x)
~^g\psi^+_{b\gamma}(y)~^g\psi_{a\varepsilon}(y)
\gamma^{(4)}_{\alpha\beta}\gamma^{(4)}_{\gamma\varepsilon} J(x,y)
\label{17} \ee where $J(x,y)$ is
\be
J(x,y) =\int^x_0 du_i\int^y_0 dv_i D(u-v),~~i=1,2,3. \label{18}
\ee Note that we have omitted everywhere for simplicity the upper
index $(Y)$
  in $\psi^{(Y)}(x)$, but it is implied, since otherwise both (\ref{14}) and
(\ref{15}) are not gauge-invariant.

\section{Bosonization of Effective Quark Lagrangians}

We shall separate out white bilinears in (\ref{17}) following the
standard procedure given for the general $L^{(n)}_{EQL}$ in
\cite{14},\cite{27},
\be
\Psi^{fg}_{\alpha\varepsilon}(x,y)\equiv ~^f\psi^+_{a\alpha}(x)
~^g\psi_{a\varepsilon}(y)= ~^f\psi^+_{a'\alpha} (x)
\Phi_{a'a}(x,Y) \Phi_{ac}(Y,y)~^g\psi_{c\varepsilon}(y) \label{19}
\ee and introduce the isospin generators $t^{(n)}_{fg}$
\be
\sum^{n^2_f-1}_{n=0}t^{(n)}_{fg}t^{(n)}_{ij}= \frac12
\delta_{fj}\delta_{gi}; t^{(0)}=\frac{1}{\sqrt{2n_f}}\hat 1.
\label{20} \ee

Hence the bilinears in (\ref{17}) can be written as
\be
\Psi^{fg}_{\alpha\varepsilon}(x,y) \Psi^{gf}_{\gamma\beta} (y,x) =
2 \sum^{n^2_f-1}_{n=0} \Psi^{(n)}_{\alpha\varepsilon}(x,y)
\Psi^{(n)}_{\gamma\beta} (y,x) \label{21} \ee where we have
defined
\be
\Psi^{(n)}_{\alpha\varepsilon}
 (x,y) \equiv ~^f\psi^+_{a\alpha}(x)t^{(n)}_{fg}~^g\psi_{a\varepsilon}(y).
\label{22} \ee Now one can use the Fierz transformation (see
Appendix 3 for more details).
\be
\gamma^{(4)}_{\alpha\beta} \gamma^{(4)}_{\alpha'\beta'}=
\frac14\sum^5_{k=1}\bar O^{(k)}_{\alpha\beta'}\bar
O^{(k)}_{\alpha'\beta} \label{23}  \ee with $$ \bar
O^{(1)}_{\alpha\beta}=\delta_{\alpha\beta},\bar O^{(2)}\bar
O^{(2)} \gamma^{(4)}\gamma^{(4)}-\gamma^{(i)}\gamma^{(i)}, \bar
O^{(3)}\bar O^{(3)}=(\gamma^5\gamma^{(4)})(\gamma^5\gamma^{(4)})-
(\gamma^5\gamma^{(i)})(\gamma^5\gamma^{(i)}), $$
\be
\bar O^{(4)}\bar
O^{(0)}=(\sigma_{ik})(\sigma_{ik})-(\sigma_{4k})(\sigma_{4k})-
(\sigma_{k4})(\sigma_{k4}) \label{24} \ee $$ \bar
O^{(5)}_{\alpha\beta}=i(\gamma^{(5)})_{\alpha\beta};~~
\sigma_{\mu\nu}=\frac{\gamma_\mu\gamma_\nu-\gamma_\nu\gamma_\mu}{2i}.
$$

Introducing (\ref{21}) and (\ref{23}) into (\ref{17}), one obtains
\be
L_{EQL}^{(2)}(el) =-\int d^4x\int
d^4y\Psi^{(n,k)}(x,y)\Psi^{(n.k)}(y,x)\tilde J (x,y) \label{25}
\ee where we have defined
\be
\Psi^{(n,k)} (x,y) =\frac12 \Psi^{(n)}_{\alpha\varepsilon}(x,y)
\bar O^{(k)}_{\varepsilon \alpha},~~\tilde J\equiv \frac{1}{N_c}
J. \label{26} \ee Bosonization is now done in a standard way using
identity (signs and indices of summation and integration are
suppressed)
\be
e^{-\Psi\tilde J\Psi}= \int(\det \tilde J)^{1/2} D\chi\exp [-\chi
\tilde J\chi+ i\Psi\tilde J \chi + i\chi \tilde J \Psi] \label{27}
\ee and hence the partition function with the only Gaussian
contribution $L^{(2)}$ assumes the form
\be
Z=\int D\psi D\psi^+ D\chi \exp L_{QML} \label{28} \ee where the
effective quark-meson Lagrangian is $$ L^{(2)}_{QML} =\int
d^4x\int
d^4y\left\{~^f\psi^+_{a\alpha}(x)[(i\hat\partial+im)_{\alpha\beta}\delta(x-y)
+iM^{(fg)}_{\alpha\beta} (x,y)]~^g\psi_{a\beta}(y)- \right.$$
\be
\left.-\chi^{(n,k)}(x,y)\tilde J(x,y) \chi^{(n,k)}(y,x)\right\}
\label{29} \ee and the effective  quark-mass operator is
\be
M^{(fg)}_{\alpha\beta}(x,y) =\sum_{n,k} \chi^{(n,k)}(x,y)\bar
O^{(k)}_{\alpha\beta}t^{(n)}_{fg}\tilde J(x,y). \label{30} \ee In
a similar way one can bosonize the term $L_{EQL}^{(3)}$. However
the computations are more lengthy and we shall present the result
in a separate publication \cite{13}.

\section{The effective chiral Lagrangian $L_{ECL} $ and stationary point analysis}

We are now in position to integrate over quark fields in
(\ref{28}) and obtain Effective Chiral Lagrangian $L_{ECL}^{(2)}$.
The result is
\be
Z=\int D\chi e^{L_{ECL}^{(2)}(\chi)} \label{31} \ee with $$
L_{ECL}^{(2)} (\chi) =- \int d^4 xd^4 y \sum_{n,k} \chi^{(n,k)}
(x,y) \tilde J (x,y) \chi^{(n,k)} (y,x)+ $$
\be
+N_c tr \ln [ (i\hat \partial +im) \delta(x-y) + i M
(x,y)].\label{32} \ee In (\ref{32}) we have taken into account
that $M$ is colorless, and the sign $tr$ refers to the summation
(integration) over Lorentz, flavor indices and space-time
coordinates.

As the next standard step one finds the stationary point equations
to determine the ground state values for the auxiliary  bosonic
fields $\chi$. Taking functional derivative of $L_{ECL}^{(2)}$
with respect to $\chi^{(n,k)}$, one obtains
\be
2\chi^{(n,k)} (x,y) \tilde J(x,y) =-i  N_c tr [ S(x,y)) \bar O
^{(k)}t^{(n)}] \tilde J (x,y) \label{33} \ee where we have defined
as in \cite{14}
\be
S(x,y)=-[(i\hat \partial +im) \hat 1 + i \hat M]^{-1}_{x,y}.
\label{34} \ee The set of (nonlinear) equations (\ref{33}),
(\ref{34}) is the central result of the present paper. In what
follows we shall study the properties of the solutions and compare
this result to the previously obtained equations in \cite{14},
where another method was used  --large $N_c$ approximation in the
Dyson-Schwinger equations for the heavy-light system with one
fixed flavour. In this case one should take $n_f=1$, and
$t^{(0)}=\frac{1}{\sqrt{2}}$. Moreover, only scalar part (since it
is dominant at large distances from the heavy source \cite{14,26}
was considered, hence one can write instead of (\ref{30})

\be
M\to M_0 (x,y) =\frac{1}{\sqrt{2}}\chi (x,y) \tilde J (x,y)
\label{35} \ee and (\ref{33}) reduces to the equation
\be
iM_0(x,y) = (\gamma_4S(x,y) \gamma_4)_{k=1} J(x,y) \label{36}\ee where we
have used the relation.
\be
(\gamma_4 S\gamma_4)_{\alpha\gamma} = \gamma^{(4)}_{\alpha\beta'}
S_{\beta'\alpha'} \gamma^{(4)}_{\alpha'\gamma} = \frac14 \sum_k
O^{(k)}_{\alpha\gamma} tr (SO^{(k)}).
 \label{37} \ee
The equation (\ref{36}) is exactly the same as Eq. (\ref{15}) in
\cite{14}  where color-electric field component is retained,
namely  in the full answer $$ iM_0(x,y) = J(x,y) \gamma_\mu S(x,y)
\gamma_\mu- J_{ik} \gamma_k S(x,y) \gamma_i \eqno{(36')}$$ one
keeps only $\mu=4,k=1$. As it is follows from the definition of $S$,
Eq.(\ref{34}), one has another equation to complete a full set
\be
(-i\hat \partial -im) S(x,y) - i\int M(x,z) S(z,y)
d^4z=\delta^{(4)}(x-y) \label{38} \ee

Let us look more closely at the scalar part of mass operator,
$M_0(x,y)$ Eqs. (\ref{35}), (\ref{36}). The properties of the
kernel $J(x,y)$ (\ref{18}) have been thoroughly investigated in
\cite{14}-\cite{16}  and it was shown there, that when $\vex$ is
close to $\vey$, then $J(x,y)$ is growing linearly at large
$|\vex|$, e.g. when for simplicity $D(u)$ is taken in the Gaussian
form
\be
D(u) =D(0) e^{-\frac{\veu^2+\veu^2_4}{4T_g^2}}\label{39}\ee then
\be
J(\vex\sim \vey, |\vex| \to \infty)= |\vex| 2T_g\sqrt{\pi}D(0)
e^{-(x_4-y_4)^2/4T_g^2}\label{40}\ee whereas \be \sigma=\frac12
\int D(u)d^2 u= 2\pi T^2_gD(0). \label{41}\ee Moreover $S(x,y)$
(\ref{34}) (where only $M_0$ is retained) at large distances
displays the properties of the smeared $\delta$ -function (see
\cite{14}-\cite{16}  for discussion and numerical estimates)
\be
\gamma_4 S(x,y) \gamma_4 \sim \tilde \delta^{(3)}
(\vex-\vey)\label{42}\ee and as  a result the product $J(x,y)
\gamma_4 S(x,y) \gamma_4$ behaves linearly in $|\vex|$ at large
$|\vex|$, in such a way that in the equation (\ref{38}) one has
\be
\int M_0(x,z) S(z,y) d^4 z\to \sigma|\vex| S(x,y).\label{43}\ee
Thus for $|\vex|\gg T_g$ one has
\be
M_0(x,z)\approx \sigma|\vex| \tilde
\delta^{(4)}(x-z)(1+O(T_g/|\vex|)\label{44}\ee and $\tilde
\delta^{(4)}(x-z)$ is smeared off at distance of the order of
$T_g$. To proceed we disregard first in the sum (\ref{30}) all
terms except for the scalar and pseudoscalar fields,
\be
M(x,y)=\chi_{S}(x,y)\frac{\tilde J(x,y)}{\sqrt{2n_{f}}}+ \hat
\chi_{\pi}(x,y) i \gamma_{5} \tilde J(x,y)\label{45}\ee with
$$\hat\chi_{\pi}(x,y)=\chi_{\pi}^{(f)} t^{f}.$$ The form
(\ref{45}) can be equivalently parametrized in a nonlinear way as
follows
\be
\hat M(x,y)=M_{S}(x,y) \hat U(x,y),\hat
U=exp(i\gamma_{5}\hat\phi), \hat \phi(x,y)=\phi^{f}(x,y) t^{f}.
\label{46}\ee Now using normalization property
\be
tr(t^{(n)} t^{(m)})=\frac{1}{2}
\delta_{nm},n,m=0,1,\ldots,n_{f}-1, \label{47}\ee one easily
obtains
\be
\frac{1}{4} tr(\hat M\hat M^+)=\frac{1}{2}(\chi^2_{s}+\ve\chi^2_{\pi})
\tilde J^{2}(x,y)=M^2_{S} n_f \label{48}\ee and hence the first term in
(\ref{32}) can be written as
\be
(\chi^2_s +\chi^2_{\pi}) \tilde J (x,y)=2n_f \tilde J^{-1}(x,y)
M^2_S(x,y) \label{49}\ee and the total Lagrangian (\ref{32}) is
\be
L^{(2)}_{ECL}(M_S,\hat \phi)=-2n_f (\tilde J (x,y))^{-1}
M^2_S(x,y)+ N_c tr\log[(i\hat\partial+im)\hat 1+iM_S \hat
U].\label{50}\ee
 The stationary point equations assume the form
\be
\frac{\delta L^{(2)}_{ECL}}{\delta M_S(x,y)}=-4n_f (\tilde J (x,y))^{-1}
M_S(x,y)-N_c tr(S i e^{i\gamma_5 \hat\phi})=0 \label{51}\ee
with
\be
 S(x,y)=-[i\hat\partial+im +iM_S \hat U]^{-1}_{x,y}, \label{52} \ee
 \be
 \frac{\delta L^{(2)}_{ECL}}{\delta \hat\phi(x,y)}=N_c tr( S
 M_S e^{i\gamma_5 \hat\phi} \gamma_5). \label{53}\ee
 The solution with $ \hat\phi=0,M_S=M^{(0)}_S$ satisfies(\ref{53})
 while(\ref{51}) may be rewritten in the form
 \be
 i M^{(0)}_{S}(x,y)=\frac{N_c}{4} tr S \tilde J(x,y)=
 N_c (\gamma_4 S \gamma_4)_{sc} \tilde J(x,y) ,\label{54}\ee where
 $(\Gamma)_{sc}$ means the scalar part of operator,as defined in
 (\ref{37}), and one can see that one recovers Eq.(\ref{36}), derived
 before in \cite{14} in a different formalism.

 \section {The derivative expansion of $ L^{(2)}_{ECL}$ }

 In this chapter chiral Lagrangian will be written as a series in
 powers of derivatives of the field $ \hat U(x,y)$. A similar procedure
 for the local case and in absence of confinement was
 systematically done in \cite{5},\cite{29}-\cite{30}, and more recently
 in \cite{31}. For the case of instanton model it was done in \cite{7}.
 In all cases at some moment the local limit of the resulting chiral
 Lagrangian is done, and confining properties of the kernel $ M_S(x,y)$
 are not taken into account. In what follows we keep both $ \hat\phi$
 and $ M_S $ nonlocal and the confining property (\ref{44}) is exploited.
 It should be noted that nonlocality of the field $\hat\phi$ is a
 necessary consequence of its quark-antiquark structure and this
 structure is lost when localization approximation is done. Since the
 radius of the pion is around $0.6$fm, it is only for small momenta that
 localization procedure is justified.
 We now turn to the second term on the r.h.s. of (\ref{50}) which
 contains pionic field and make an expansion of its real part in powers
 of derivatives of the field $ U $. Note that imaginary part of
 effective chiral Lagrangian was studied in \cite{5},\cite{7},\cite{30}
 and it starts with the terms of the 5-th power in pionic field; it will
 not be studied below. Defining the real part of the pionic effective
 action $Re L_{eff}[\pi]$, one has
 \be
 Re L_{eff}[\pi]=- \frac{N_{c}}{2}\log det
 \frac{D^{+}D}{D^{+}_{0} D_{0}}, \label{55} \ee where notations are used
 \be
 D=i\hat\partial +im +i M_{S} \hat U,D_{0}=i\hat\partial+
 im +i M_{S}. \label{56} \ee Moreover
 \be
 D^{+}D =- \partial^{2}-\partial_{\mu}( M_{S} \hat U^{+})\gamma_{\mu}+
  (m+M_{S})^2 +m M_{S}(\hat U^{+}+\hat U) \label{57} \ee
 \be
 D^{+}_{0} D_{0}=-\partial^{2} + (m+M_{S})^{2}. \label{58} \ee
 To simplify we put $  m=0 $ and rewrite (\ref{55}) as
 \be
 Re L_{eff}[\pi]= - \frac{N_{c}}{2} tr \log ( 1 - G_{0} \partial_{\mu}
 M_{S} \hat U^{+} \gamma_{\mu}), \label{59} \ee where we have defined
 \be
 G_{0}(x,y)=(-\partial^{2} + M^{2}_{S})^{-1}_{x,y}. \label{60} \ee

 In (\ref{59})the sign $tr$ implies the sum over Lorentz and flavour
 indices and integral over coordinates, having in mind that every
 factor appearing in (\ref{59}) under the logarithm sign should be
 considered as a matrix in the space-time coordinates.
 Expanding logarithm in (\ref{59}) and keeping only 2nd and 4th
 order terms in $\partial_{\mu} \hat U^{+}$, one obtains
 $$
 Re L_{eff}[\pi]= N_{c} \overline{tr} Re [G_{0}\partial_{\mu} U^{+}
 G_{0}\partial_{\mu} U
 $$
 \be
  + \frac{1}{2} G_{0} \partial_{\mu} U^{+}
 G_{0}\partial_{\nu} U G_{0} \partial_{\alpha} U^{+} G_{0}\partial_{\beta}
 U (\delta_{\mu\nu}\delta_{\alpha\beta} +\delta_{\mu\beta} \delta_{\nu\alpha}
 -\delta_{\mu\alpha} \delta_{\nu\beta})] + \ldots \label{61}\ee
 Here $\overline{tr}$ implies summing over flavour indices and integration
 over coordinates, and $ U=M_{S}(x,y) e^{i\hat\phi(x,y)}$.
 We consider now the first term on the r.h.s. of (\ref{61}) and write
 explicitly the coordinate part of $\overline{tr}$
 \be
 Re L^{(2)}_{eff}[\pi]= N_{c} tr_{f} Re [\int d^{4}x G_{0}(x,y)
 \partial_{\mu} U^{+}(y,z) d^{4}z G_{0}(z,u) d^{4}u \partial_{\mu}
 U(u,x)]. \label{62} \ee
 In the limit, when $M_{S}(x,y)$ becomes local,
 $$ U(x,y)\approx \tilde
 \delta^{4}(x-y) M_{S}(x) e^{I\hat\phi(x)}=\tilde\delta^{4}(x-y) U(x),$$
 one has
 \be
 Re L^{2}_{eff}[\pi]= N_{c} tr_{f} Re[\int d^{4}x d^{4}y G_{0}(x,y)
 \partial_{\mu} U^{+}(y) G_{0}(y,x) \partial_{\mu} U(x)]. \label{63}\ee
 One can see that Eq.(\ref{63}) yields a nonlocal chiral Lagrangian,
 the nonlocality being given by the range of the quark Green's function
 $G_{0}(x,y)$. In case of a model without confinement, e.g.instanton
 model or NJL model, $G_{0}(x,y)$ would represent the free Green's
 function of a massive quark,with the constituent mass $\mu$ created
 by chiral symmetry breaking, $\mu \approx 0.3 GeV$. Hence the range of
 nonlocality in this model case is large,and the derivative expansion
 in powers of $\partial U$, which is a standard systematic procedure in
 \cite{5},\cite{7},\cite{30} is justified for small momenta $p \leq \mu$.
 At this point one should remember that in (\ref{61})-(\ref{63}) both
 $M_{S}(x,y)$ and $U(x,y)$ (and hence $S(x,y)$, Eq(\ref{34})), are
 defined gauge-invariantly with respect to the contour $Y$, which for
 simplicity was taken to be $x_{4}$ axis. Therefore the resulting string
 in $\hat M,M_{S}$ goes from the points $(x,y)$ (coinciding when $T_{g}$
 tends to zero) to the contour $Y$. Physically it means,that the effective
 Lagrangian (\ref{62}) describes the bosonic field (e.g.pion) in the
 presence of the heavy quark. It will enable us to define in the following
  chapter matrix elements of heavy-light meson transitions with
  emission of a pion.
  The case of effective boson Lagrangian for light quarks and antiquarks
  (i.e. without the heavy quark line) will be considered in the second
  paper of this series.

  \section{Pionic transitions in heavy-light mesons}

  In this section the main emphasis will be on the pionic part of the
  quark mass operator, which enters the quark-meson Lagrangian as
  \be
  \Delta L=i\int d^{4}x d^{4}y\psi^{+}(x)\hat M(x,y)\psi(y)\label{64}\ee
 where $\hat M(x,y)$ according to (\ref{46}) can be written in the form
 \be
 \hat M(x,y)= M_{S}(x,y) e^{i\gamma_{5}\hat\phi(x,y)}. \label{65}\ee
 In the limit of small $T_{g}$ one obtains for $M_{S}(x,y)$ a localized
 expression
 \be
 M_{S}(x,y) \approx \sigma|\vex| \delta^{(4)}(x-y),|\vex|\gg T_{g} \label{66}
 \ee and the corresponding linearized in $\hat\phi$ Lagrangian is (in the
 Minkowskian space-time)
 \be
 \Delta L^{(1)}= \int \overline{\psi}(x)\sigma |\vex|\gamma_{5}
 \frac{\pi^{A}\lambda^{A}}{F_{\pi}} \psi(x) dt d^{3}x \label{67} \ee
 where $F_{\pi}$ is known \cite{3} to be $F_{\pi}=94 MeV$.
 For the pionic transition between heavy-light states one should compute
 matrix elements
 \be
 \lan M_{2}(\vep_{2}),\pi(\vek) | \Delta L^{(1)}|M_{1}(\vep_{1})\ran .\label{68}
 \ee
 Neglecting recoil momentum of heavy-light meson one effectively
 reduces the problem to the calculation of the matrix element
 \be
 W_{21} =\int \overline{\psi}_{2}(x)\frac{\sigma |\vex|}{F_{\pi}}
 \gamma_{5} \frac{\lambda^{A} e^{i \vek \vex}}{\sqrt{2 \omega_{\pi}(\vek) V}}
 \psi_{1}(\vex) d^{3} x \label{69} \ee and the decay probability is
 \be
 w = 2 \pi | W_{21}|^{2} \delta (E_{1}-E_{2}-\omega) \frac{V d^{3}k}{(2\pi)^{3}}. \label{70}
  \ee
 In (\ref{66}) $\overline{\psi}_{2}(x),\psi_{1}(x)$ are quark wave
 functions  of the heavy-light states 2 and 1 respectively,which can
  be taken from the solutions of the corresponding Dirac equations,
  found in \cite{32},\cite{26}. At this point one can rewrite the matrix
  element of $\Delta L^{(1)}$ between two stationary quark states
  $\psi_{m},\psi_{n}$. Using Dirac equations
  \be
  [\veal \vep + \beta (m + \sigma |\vex|) + V_{Coul}] \psi_{n}=
  \epsilon_{n} \psi_{n} \label{71} \ee
  \be
  \overline{\psi}_{m}[-\veal \vep + \beta (m +\sigma |\vex|)+
 V_{Coul}]=\epsilon_{m} \overline{\psi}_{m} \label{72} \ee
  one obtains
  \be
\lan m |\Delta L^{(1)} |n\ran =\frac{1}{2 F_{\pi}} \lan m | - 2m
\gamma_{5} \hat\pi+
  \beta \gamma_{5}(\epsilon_{m}-\epsilon_{n}) \hat\pi +
  \gamma_{5} \beta \veal \vep \hat\pi |n\ran .\label{73} \ee
  In the chiral limit, $m=0$, one can rewrite the last two terms inside
  the brackets in (\ref{73}) as
  \be
  \gamma_{5} (\vegam \frac{\partial \hat\pi}{\partial\vex} +
  i \beta \frac{\partial \hat\pi}{\partial t}) = \gamma_{5}
  \gamma_{\mu} \partial_{\mu} \hat\pi. \label{74} \ee
 One obtains from (\ref{73}) the form of the quark-pion interaction
 which one usually writes as (see \cite{33},\cite{34} and refs. therein)
 \be
 \Delta L^{ch} = g^{q}_{A} tr (\overline\psi\gamma_{\mu} \gamma_{5}
 \omega_{\mu} \psi),\omega =\frac{i}{2F_{\pi}}(u \partial_\mu u^{+}-
 u^{+} \partial_{\mu} u), \label{75} \ee
 where $u=\sqrt{\hat U}$ and
$g^{q}_{A}$ is the axial vector coupling of the (constituent)
quark. It is easy to see that (\ref{73}) is the first term in the
expansion of (\ref{75}) in powers of  the field $\hat\pi$, and it
is gratifying that the parameter $g^{q}_{A}$ is defined
theoretically to be equal exactly to 1 in our local
approximation,which is in agreement with large $N_{c}$ argument in
\cite{33}.

\section{Concluding remarks}

We have performed a systematic bosonization procedure starting
from the QCD Lagrangian and have derived the nonlocal chiral
Lagrangian and its limiting local form for quarks interacting with
pionic field. By the method of construction, the resulting
Lagrangian is applicable for quark-pion interaction when the quark
is coupled by the string to the heavy antiquark. Therefore our
results can be immediately applied to the pionic transitions in
the heavy-light mesons. The one-pion transitions have been studied
using the form (\ref{75}) in \cite{34}, and it was found from
comparison to experiment that the values of $g^{q}_{A}$ around 0.7
are preferred. It makes it reasonable to study the nonlocal
version of the chiral Lagrangian (\ref{73}), since nonlocality
effectively decreases resulting matrix elements, possibly
explaining the mismatch between our theoretical value
$g^{q}_{A}=1$ and observed value of 0.7. In a similar way one can
obtain matrix elements for double pion emission in heavy boson
transitions, which are of special interest for $(B,D),(B,D*)$
semileptonic decays. The case of purely light mesons, can be
treated  in a similar way, but needs another string configurations
to be taken into account. This is planned to do in the subsequent
paper of this series. The author is grateful to F.Gross and
I.Musatov for discussions and to J.Goity for useful criticism and
suggestions. This work was supported by DOE contract
DE-AC05-84ER40150 under which SURA operates the Thomas Jefferson
National Accelerator Facility.

\section*{Appendix 1}

\subsection*{Vacuum averages of Field Correlators}

\setcounter{equation}{0} \def\theequation{A1.\arabic{equation}}

 \vspace{1cm}

For the field operator transported to the point $Y$,
\be
F_{\mu\nu}^{(Y)}(x) = \Phi(Y,x) F_{\mu\nu} (x)
\Phi(x,Y)\label{A1.1} \ee so that gauge transformation has the
form
\be
F_{\mu\nu}^{(Y)} (x)\to U^+ (Y)F_{\mu\nu}^{(Y)}(x)
U(Y),\label{A1.2}\ee the vacuum average of any product can be
expressed through Kronecker symbols, e.g.
\be
g^2\lan
(F^{(Y)}(x))_{ab}(F^{(Y)}(y))_{cd}\ran=\delta_{ad}\delta_{bc}
\mathcal{P}(x,y)\label{A1.3}\ee
 where $ab,cd$ are fundamental color indices and
 $\mathcal{P}(x,y)$ can be easily connected to the color trace,
 i.e. to the FC,
 \be
 \mathcal{P}(x,y) = \frac{\delta_{ad}\delta_{bc}}{N_c^2}
  g^2 \lan tr ( F^{(Y)}(x) F^{(Y)}(y))\ran.\label{A1.4}\ee

Similar rules apply to products of any number of $F^{(Y)}$, they
can be deduced from the global $SU(N_c)$ rules for the tensors
averaged with the Haar measure, namely one can make a gauge
 transformation
\be
(F^Y_{\mu\nu} (x))_{ab}\to
\Omega^+(Y)_{ac}(F^Y_{\mu\nu}(x))_{cd}\Omega(Y)_{db}\label{A1.5}\ee
and average over $D\Omega$ with the usual Haar measure. In this
way one obtains relation (\ref{8}). At this point it is
interesting to note that the product of three $F$'s may have
another representation in the $SU(3)$ group, since there one can
use totally antisymmetric tensor $e_{abc}$.

E.g. one can write the following equality $$e_{ace} e_{bdf} =
\delta_{ab} (\delta_{cd} \delta_{ef}-\delta_{cf} \delta_{ed}+ $$
$$ + \delta_{ad} (\delta_{cf}\delta_{eb} -\delta_{cb}
\delta_{ef})+ $$
\be
+ \delta_{af}(\delta_{cb } \delta_{ed} -\delta_{cd} \delta_{eb})
\label{A1.6} \ee

Therefore on can use two forms of writing for the product. In the
first case one writes $$ \lan (F^{(Y)}(x))_{ab}
(F^{(Y)}(y))_{cd}(F^{(Y)}(z))_{ef}\ran = \mathcal{P}_1(x,y,z)
\delta_{bc} \delta_{de}\delta_{af}+ $$
\be
\mathcal{P}_2(x,y,z) \delta_{bc} \delta_{fc}
\delta_{ad}\label{A1.7}\ee and finds $ \mathcal{P}_1,
\mathcal{P}_2$ by multiplying both sides of (\ref{A1.7}) with the
corresponding contribution of $\delta$-symbols. In this way one
arreves at the  Eq. (\ref{8}) in the main text. If instead one
uses (\ref{A1.6})  instead of one of  combinations, or separates
(\ref{A1.6}) out of  Eq. (\ref{8}), one arrives at the white
$(3q),(3\bar q)$ combinations.

\section*{Appendix 2}

\subsection*{Vacuum averages of $\lan (A)^n\ran$}

\setcounter{equation}{0} \def\theequation{A2.\arabic{equation}}

 \vspace{1cm}

 One can use the generalized contour gauge expression for
 $A_\mu(x)$.
 \be
 A_{\mu}(x) =\int^x_Y dz_\nu\frac{\partial z_\rho}{\partial
 x_\mu} F_{\nu\rho}(z) \equiv \int^x_Y d\Gamma_{\mu\nu\rho} (z)
 F_{\nu\rho} (z)
 \label{A2.1}\ee
 to represent the average of the  product of any number of
 operators $A_\mu$ as
 $$ \lan (A_{\mu_1}^{(x_1)} )_{a_1b_1} ... (A_{\mu_n}
 (x_n))_{a_nb_n}\ran=$$
 \be
 =\int^{x_1}_Y d\Gamma_{\mu_1\nu_1\rho_1} (z_1) ...
\int^{x_n}_Y d\Gamma_{\mu_n\nu_n\rho_n} (z_n) \lan
(F^{(Y)}_{\nu_1\rho_1}(z_1))_{a_1 b_1}...
(F^{(Y)}_{\nu_n\rho_n}(z_n))_{a_n b_n}\ran. \label{A2.2} \ee

In a particular case of the modified Fock-Schwinger gauge one has:
$Y=0, x_i\to \vex_i$
\be
d\Gamma_{\mu\nu\rho} (z) = \alpha_\mu(z) dz_\nu \delta_{\rho\mu}
\label{A2.3}\ee

\section*{Appendix 3}

\subsection*{Fierz transformations}

\setcounter{equation}{0} \def\theequation{A3.\arabic{equation}}

 \vspace{1cm}

 In this appendix the derivation is given of Fierz tranformations
 for combinations of $\gamma_\mu$ matrices met in the text above.
 It is based on the clear presentation done in the book \cite{29}.
 Note however that we are always working with the Euclidean
 $\gamma$ matrices, therefore some details and coefficients
 obtained below are different from \cite{29}.

 We start with the general expansion for any $4\times 4$ matrix
 \be
 \gamma=\frac14 \sum_A C_A\gamma_A,~~ A=1,... 16,
 \label{A3.1}\ee
 \be
\left. \begin{array}{lll}
 \gamma_A=1,& A=1&\\
 \gamma_A=\gamma_\mu,& \mu=1,2,3,4; & A=2,3,4,5\\
 \gamma_A=\sigma_{\mu\nu}=\frac{\gamma_\mu\gamma_\nu-\gamma_\nu\gamma_\mu}{2i},&&A=
 6,7,8,9,10,11\\
 \gamma_A=\gamma_5\gamma_\mu,&& A=12,13,14,15\\
 \gamma_A=\gamma_5&& A=16
 \end{array} \right\}
\label{A3.2}\ee

In (A3.1)  one can derive the general representation for any
matrices $F_{mk}, G_{il}$, indeed from relation
$$\frac14\sum_A\Delta_A\gamma^A_{ml} \gamma^A_{ik}
=\delta_{mk}\delta{il},$$ changing $m\to m'$, $l\to l'$ and
multiplying with $F_{mm'} G_{l'l}$ one obtains
\be
F_{mk} G_{il} =\frac14 \sum_A\Delta_A (F\gamma_AG)_{ml}
(\gamma_A)_{ik}. \label{A3.3} \ee

For Euclidean matrices  $\gamma_A $ one easily obtains
$\Delta_a=-1$, for $A=12,13,14,15$ and $\Delta_A=1$ otherwise.
Taking $F=G=\gamma^4$, one obtains $$ \gamma^4_{mk}\gamma^4_{il}
=\frac14 \sum_A\Delta_A(\gamma^4 \gamma_A\gamma^4)_{ml}
(\gamma_A)_{ik}= $$
$$=\frac14\{(1)(1)+(\gamma^4)(\gamma^4)-(\gamma^i)(\gamma^i)+(\sigma^{ik})(\sigma^{ik})
-(\sigma^{4k})(\sigma^{4k})- $$ $$-(\sigma^{k4})(\sigma^{k4})+
(\gamma^5\gamma^4)
(\gamma^5\gamma^4)-(\gamma^5\gamma^i)(\gamma^5\gamma^i)-(\gamma^5)(\gamma^5)\}=
$$ \be\frac14\sum^5_{k=1}\bar O^{(k)}_{ml}\bar
O^{(k)}_{ik}.\label{A3.4} \ee

The last relation coincides with (\ref{23}) and operators $\bar
O^{(k)}$ are given in (\ref{24}). Note that in the curly brackets
in (\ref{A3.4}) each product of $\gamma$ matrices,
$(\gamma\gamma)(\gamma\gamma)$, has the
same order of indices as in $\bar O^{(k)}_{ml}\bar O^{(k)}_{ik}$.
In a similar way one can represent the combination of  spacial
$\gamma$ matrices, $n=1,2,3,$ no summation over $n$,
$$\gamma^n_{mk}\gamma^n_{il} =\frac14
\{(1)(1)+(\gamma^4)(\gamma^4)-(\gamma^m)(\gamma^m)+(\gamma^n)(\gamma^n)+
$$ $$+(\sigma^{\mu\nu})(\sigma^{\mu\nu})_{n\neq\mu\nu}-
(\sigma^{n\nu})(\sigma^{n\nu})-(\sigma^{\nu n})(\sigma^{\nu n})-
(\gamma^5\gamma^\mu) (\gamma^5\gamma^\mu)_{\mu\neq n}+ $$
\be
+(\gamma^5\gamma^n)(\gamma^5\gamma^n)-(\gamma^5)(\gamma^5)\}_{ml,ik}
.\label{A3.5} \ee

Summing (\ref{A3.5}) over $n$ and adding (\ref{A3.4}) one obtains
\be
(\gamma^\mu)_{mk}(\gamma^\mu)_{il}=\{(1)(1)-\frac12
(\gamma^\mu)(\gamma^\mu)
-\frac12(\gamma^5\gamma^\mu)(\gamma^5\gamma^\mu)-(\gamma^5)(\gamma^5)\}_{ml,ik}.
\label{A3.6}\ee Now in case of generalized contour gauge as in
Appendix 2 one has to make Fierz transformation of the combination
\be
(\gamma^\mu)_{mk}(\gamma^\nu)_{il}= \frac14
\sum_A\Delta_A(\gamma^\mu \gamma_A\gamma^\nu)_{ml}
(\gamma_A)_{ik}.\label{A3.7}\ee

Note that scalar and pseudoscalar combinations occur on the l.h.s.
of (\ref{A3.7}) only for $\mu=\nu$.

\end{document}